\documentstyle[prl,aps,floats,psfig]{revtex}

\begin{document}

\def\avg#1{\left< #1 \right>}
\def\abs#1{\left| #1 \right|}
\def\recip#1{\frac{1}{#1}}
\def\vhat#1{\hat{{\bf #1}}}
\def\recip#1{\frac{1}{#1}}
\def\smallfrac#1#2{{\textstyle\frac{#1}{#2}}}
\def\smallrecip#1{\smallfrac{1}{#1}}

\draft % needed to print out suggested PACS numbers

% Prevent a float page unless the situation is extreme
\def\floatpagefraction{.8}

\twocolumn

\title{Symmetry breaking in crossed magnetic and electric fields}
\author{C.~Neumann, R.~Ubert, S.~Freund, E.~Fl\"othmann,
  B.~Sheehy,\cite{Sheehy-addr} and K.~H. Welge\cite{correspondence}}
\address{Fakult\"at f\"ur Physik, Universit\"at Bielefeld,
  Universit\"atsstra\ss{}e 25, D-33615 Bielefeld, Federal Republic of
  Germany}
\author{M.~R. Haggerty and J.~B. Delos\cite{correspondence}}
\address{Physics Department, College of William and Mary,
  Williamsburg, Virginia 23187}
\date{\today}

\maketitle

\begin{abstract}
  % Abstract is limited to 600 characters.
  %\hsize\textwidth \leftskip=0.10753\textwidth \rightskip\leftskip
  %\unskip %
  We present the first observations of cylindrical symmetry breaking
  in highly excited diamagnetic hydrogen with a small crossed electric
  field, and we give a semiclassical interpretation of this effect.
  As the small perpendicular electric field is added, the recurrence
  strengths of closed orbits decrease smoothly to a minimum, and
  revive again.  This phenomenon, caused by interference among the
  electron waves that return to the nucleus, can be computed from the
  azimuthal dependence of the classical closed orbits.
\end{abstract}
\pacs{32.60.+i, 03.65.Sq, 05.45.+b}

In 1969, Garton and Tomkins reported what they called ``quasi-Landau
resonances''---oscillatory modulations in the near-ionization
absorption spectrum of barium in a strong magnetic field
\cite{GartonTomkins}.  Edmonds analyzed these oscillations
semiclassically by considering the motion of the Rydberg electron in
the plane perpendicular to the magnetic field \cite{Edmonds}.  In this
plane there is a simple family of {\em closed orbits\/}: classical
electron orbits that start at the ionic core and, moving under the
combined effect of the Coulomb and magnetic fields, eventually return
to the core.  He found that Bohr-Sommerfeld quantization of these
orbits predicts energy level spacings equal to the period of the
modulations seen in the experiment.  Later experiments have
demonstrated that the absorption spectra of highly-excited atoms in
various field configurations are influenced by many two- and
three-dimensional closed orbits
\cite{magnetic,electric,parallel,crossed}.

According to {\em closed orbit theory\/} \cite{DuDelos,Bogomolny},
each classical orbit that starts at the atom carries along with it a
portion of the outgoing electron wave.  If such an orbit returns to
the nucleus, its companion wave interferes with the steady stream of
outgoing waves, and this interference produces an oscillation in the
absorption spectrum.  The return of the electron, and the resulting
interference oscillation in the spectrum, is called a {\em
  recurrence}.  The period of the oscillation is inversely
proportional to the classical action of the closed orbit, and its
strength is proportional to the electron current density in the
neighborhood of that orbit.  Thus, the more that orbits are focused on
the atom, the stronger the associated oscillation in the spectrum,
i.e., the stronger the recurrence.

Most of the experiments and theory to date have dealt with Rydberg
atoms in static external fields along a single axis---magnetic
\cite{magnetic}, electric \cite{electric}, or parallel electric and
magnetic fields \cite{parallel}.  In such cylindrically symmetric
systems, the angular momentum component $L_z$ (i.e., $p_\phi$) along
the field axis ($\vhat{z}$) is conserved.  Classical orbits are
therefore organized into cylindrical families, with orbits within a
family differing from one another only by a rotation about the $z$
axis.  When orbits in a family return to the core, they do so in a
perfect circular focus, giving a strong recurrence.

The dynamics is profoundly changed when the cylindrical symmetry is
broken by crossing the magnetic field with an electric field.  Now
instead of a continuous family of classical orbits returning in
perfect focus, only two isolated orbits return exactly to the origin.
Meanwhile, the nonconservation of $p_\phi$ makes it impossible to
separate the $\phi$ motion \cite{3D-Divergence}.  There have been
several studies of recurrences in crossed fields \cite{crossed}, but
none of them address the present issue: the {\em transition\/} from
symmetry to asymmetry; i.e., from two to three nonseparable degrees of
freedom.  How the dynamics behaves under such symmetry change and how
it is described in semiclassical theory is a basic problem
\cite{Creagh}.  What is the signature of symmetry breaking in an
atomic absorption spectrum?

In this letter we report the first experiments on the dynamical
evolution of Rydberg hydrogen atoms in a strong magnetic field when
the cylindrical symmetry is broken by adding a small crossed electric
field.  We present results obtained for the Garton-Tomkins-Edmonds
(GTE) recurrence, which shows the characteristic signature of symmetry
breaking.  (We have observed the same signature for other orbits
\cite{further-work}.)  We also present a quantitative semiclassical
theory capable of describing this symmetry breaking, and use it to
calculate the evolution of the recurrence strength.

\paragraph{Experiment.}

The experiments are performed, as previously \cite{ExpSVS}, with
deuterium atoms in a beam collimated along the magnetic field axis,
excited at the center of the crossed fields by pulsed lasers
perpendicular to the atomic beam.  The excitation occurs in two steps,
$\text{D}(1s) + h\nu_1 \rightarrow \text{D}(2p) + h\nu_2 \rightarrow
\text{D}^*(n,m_\ell=0)$, to energies near the field-free ionization
limit ($n \approx 62\text{--}112$).  $\text{D}^*$ atoms in long-lived
states, proceeding at thermal velocity, are detected via electrons
created by field ionization.  Electrons from atoms promptly ionized
within the interaction region are prevented from crossed-field
drifting by turning off the crossed electric field quickly after the
excitation, and are trapped there temporarily then released to the
detector.  The combined signal of the two electron groups is thus
proportional to the total absorption cross section
\cite{further-work}.

The Hamiltonian for hydrogen in crossed fields (in atomic units,
$\hbar = e = m_e = 1$) is
\begin{equation}
  H = \smallrecip{2} p^2
      - 1/r
      + \smallrecip{2} B L_z
      + \smallrecip{8} B^2 (x^2 + y^2)
      + F x .
\label{eq:Hamiltonian}
\end{equation}
Due to a classical scaling law for this system, the shapes of
classical orbits do not depend on the energy $E$, electric field
strength $F$, and magnetic field strength $B$ separately, but only on
the scaled energy $\epsilon \equiv E B^{-2/3}$ and scaled electric
field $f \equiv F B^{-4/3}$.  The classical action $S \equiv \int {\bf
  p} \cdot d{\bf q}$ scales as $S = \tilde{S} w$, where $\tilde{S}$ is
a scaled action that also only depends on $\epsilon$ and $f$, and $w
\equiv B^{-1/3}$.  We use the method of {\em scaled variable
  recurrence spectroscopy\/} \cite{ExpSVS}: we measure the
photoabsorption cross section $\sigma$ as a function of $w$ while
varying the laser frequency and external fields simultaneously in such
a way as to keep $\epsilon$ and $f$ constant.  According to closed
orbit theory, such a spectrum is a smooth background plus a sum of
sinusoidal oscillations,
\begin{equation}
  \sigma(w) = C_0 + \sum_k C_k \sin \left( \tilde{S}_k w
      - \gamma_k \right) ,
\label{eq:absorption}
\end{equation}
where $C_0$ is the absorption in the absence of recurrences; the sum
is over classical closed orbits; $C_k$ is the recurrence amplitude of
orbit $k$; and $\gamma_k$ is an additional phase related to the Maslov
index.  All of the quantities in Eq.~(\ref{eq:absorption}) are
constant or slowly-varying across a scaled spectrum, except for $w$.
Thus, the absolute-square of the Fourier transform of a scaled
spectrum with respect to $w$ produces a {\em recurrence spectrum},
which has a peak of height $\abs{C_k}^2$ (the ``recurrence strength'')
at the scaled action $\tilde{S}_k$ of each classical closed orbit.

\begin{figure}
  \centerline{\psfig{figure=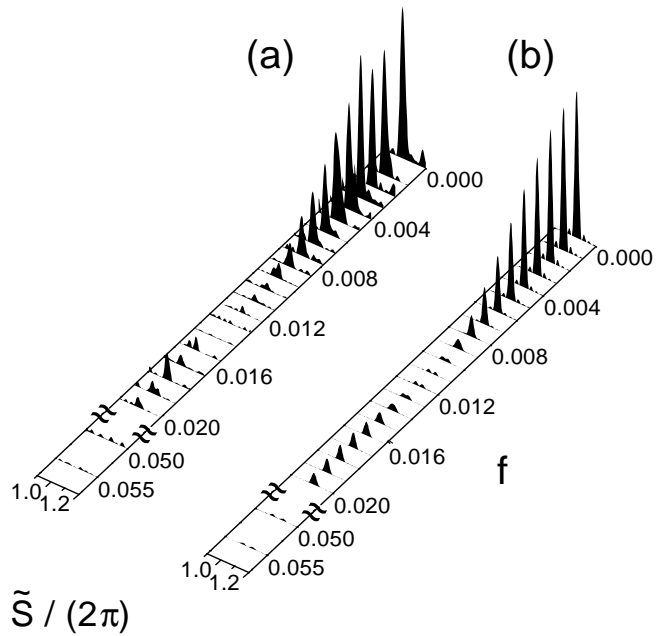,width=8.5cm,angle=270}}
  \vspace{4.0mm}
  \caption[Short title]{
    These segments of (a) experimental and (b) theoretical scaled
    action spectra show how cylindrical symmetry breaking affects the
    strength of the Garton-Tomkins-Edmonds recurrence.  The data are
    for deuterium at constant scaled energy $\epsilon = -0.15$, where
    the classical dynamics is fully chaotic.  $B$ was scanned from
    $1\text{ T}$ to $6\text{ T}$, i.e., $61.7 \geq w \geq 34.0$.  The
    GTE recurrence peak is shown for several values of the (scaled)
    strength of the perpendicular electric field (the vertical scale
    is arbitrary).  The recurrence is strongest when $f=0$, where it
    is enhanced by perfect focusing due to cylindrical symmetry.  As
    the electric field is increased, the strength falls off quickly,
    vanishes, then increases again in a weak revival about 10\% of the
    original peak height.  (Note that there is a break in the $f$
    axis.)  }
  \label{fig:experiment}
  \label{fig:theory}
\end{figure}

The GTE recurrence reported in this paper, the lowest-action peak in
these recurrence spectra, results from the family of ``rotator''-type
orbits that lie in the $z=0$ plane.  Figure~\ref{fig:experiment}(a)
shows this recurrence peak for a range of scaled electric fields,
exhibiting the essential signatures of symmetry breaking and the
transition from 2 to 3 non-separable degrees of freedom.
The falloff and revival behavior, qualitatively observed also for other
rotator and vibrator type orbits, can be explained by a semiclassical
theory of symmetry breaking.

\paragraph{Classical and semiclassical symmetry breaking.}

\begin{figure*}
  \centerline{
    \psfig{figure=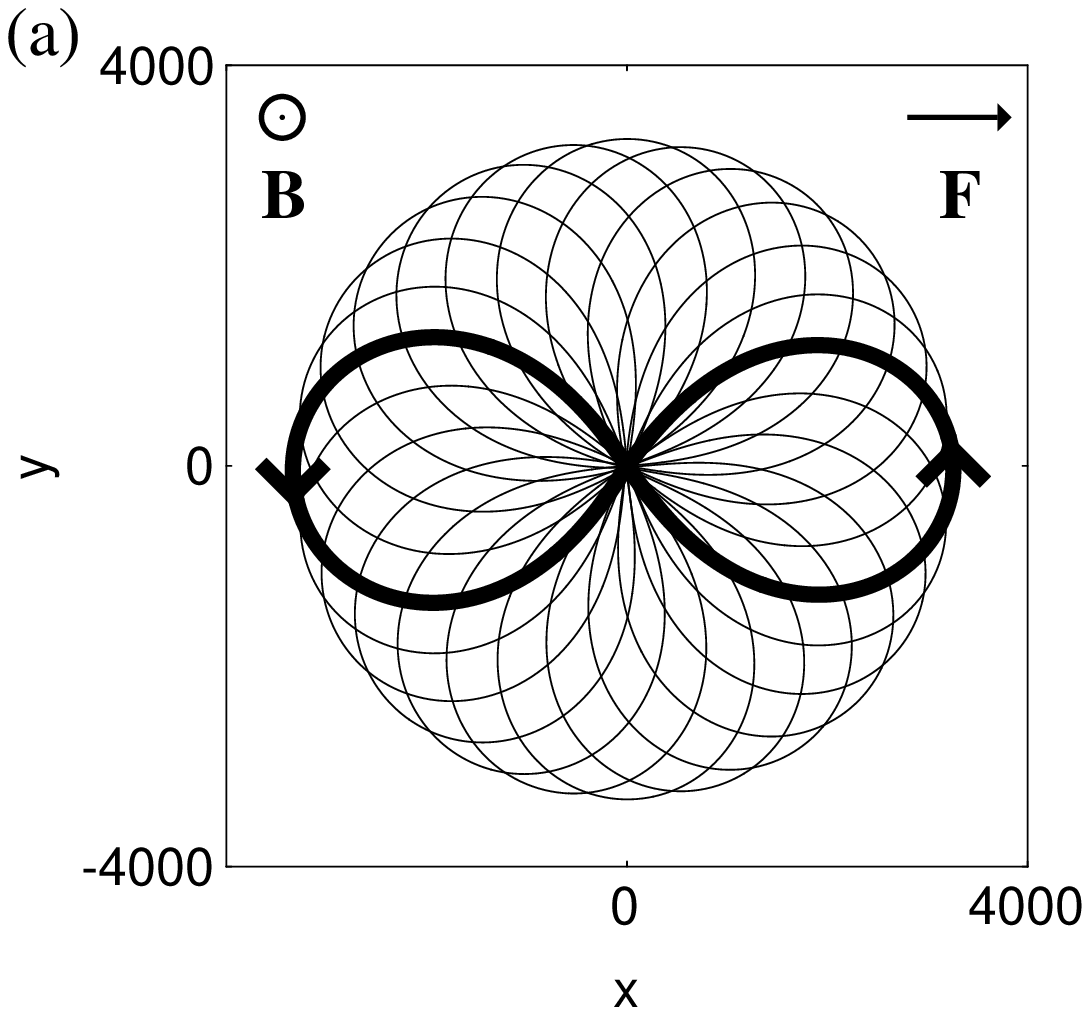,width=5.3cm}
    \psfig{figure=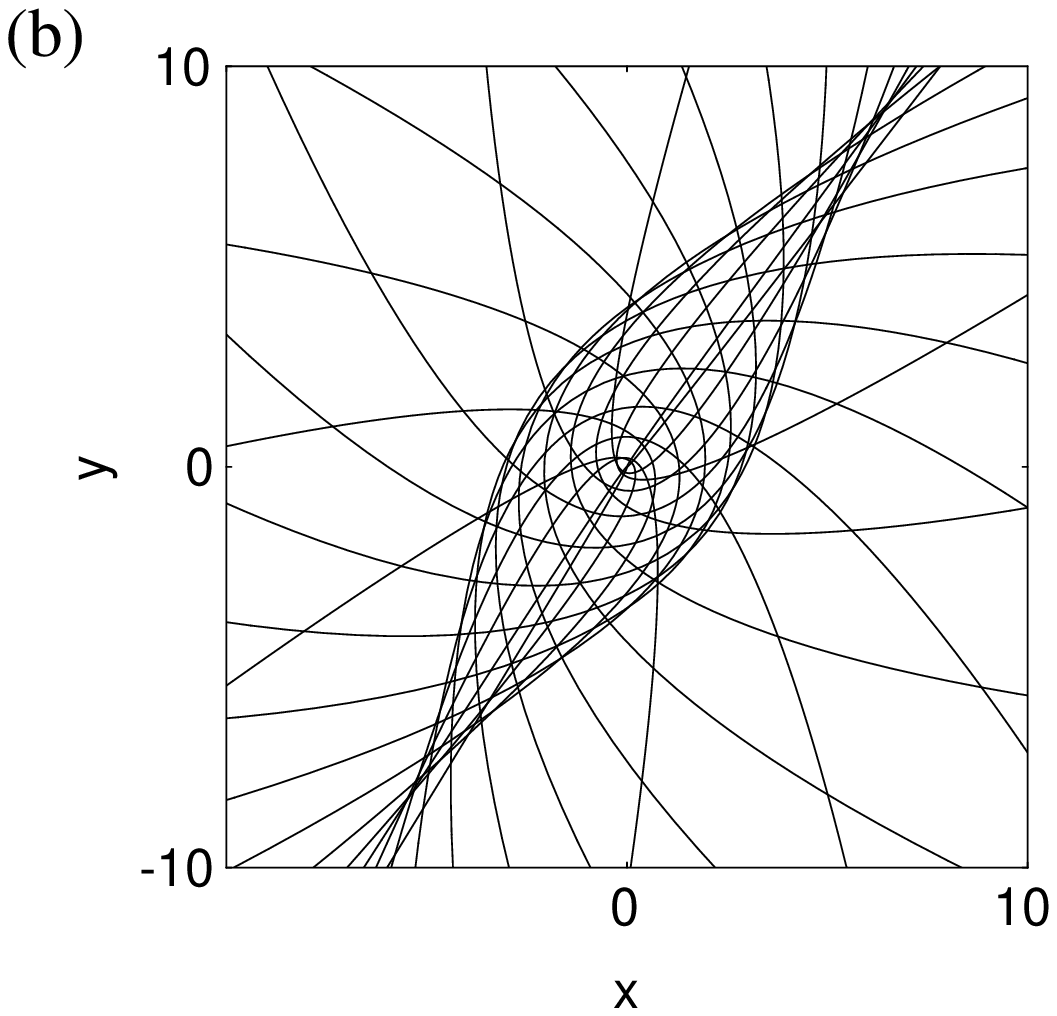,width=5.3cm}
    \psfig{figure=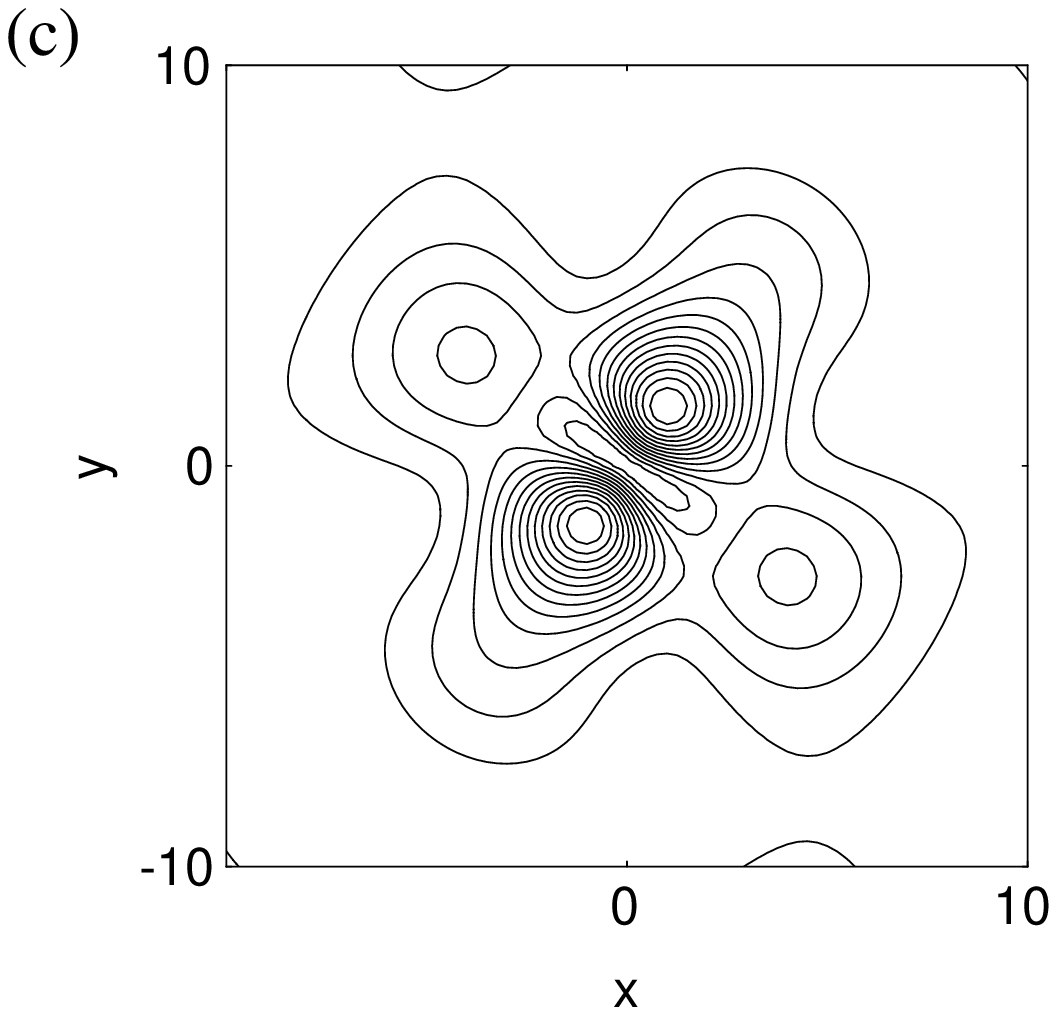,width=5.3cm}
  }
  \caption[Short title]{
    Even a slight perpendicular electric field destroys the classical
    cylindrical symmetry.  These figures were computed for the
    parameters $\epsilon=-0.15$, $f=0.012$, and $B=3\text{ T}$;
    distances are measured in Bohr radii.  (a) The family of electron
    orbits associated with the GTE recurrence.  These orbits lie in
    the $(x,y)$ plane, perpendicular to ${\bf B}$.  For nonzero
    electric field, only two orbits (shown with heavy lines) return
    exactly to the nucleus.  (b) A closeup of the returning part of
    the trajectories as they approach the nucleus.  The two closed
    orbits return to the nucleus diagonally; all others follow
    near-parabolic paths near the nucleus.  (c) The symmetry-broken
    returning wavefunction $\abs{\psi}^2$ associated with the same
    family of orbits, computed using the actions of
    Eq.~(\ref{eq:S-dependence}).  The wavefunction is generally large
    where the trajectories are dense, but destructive interference at
    the nucleus leads to a weak recurrence amplitude at this choice of
    parameters.  }
  \label{fig:miss}
  \label{fig:quantum-cusp}
\end{figure*}

As explained in the introduction, the perpendicular electric field
destroys the cylindrical symmetry of hydrogen in a magnetic field.  It
can be shown that from each family $k$ of formerly-closed orbits, just
two isolated orbits return to the origin when $f \neq 0$ (see
Figs.~\ref{fig:miss}(a) and (b)).
From a classical perspective, therefore, one might expect a sudden
drop in the recurrence strength when the electric field is turned on.
However, electron waves, unlike classical trajectories, diffract.
Therefore, the whole family of returning trajectories, even those that
miss the atom, still contributes to the recurrence.  But waves
arriving from different directions have different phases, so their
interference is partially-destructive, and the strength of the
recurrence is reduced.

\begin{figure}
  \centerline{\psfig{figure=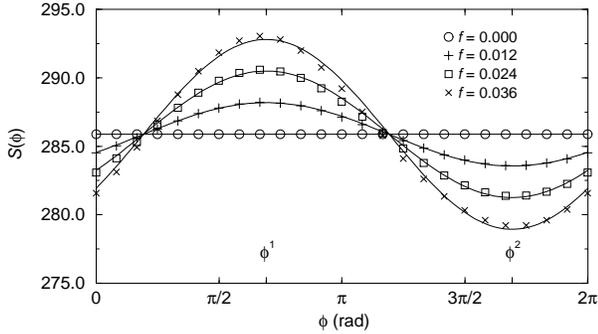,width=8.0cm}}
  \vspace{4.0mm}
  \caption[Short title]{
    Action $S(\phi)$ vs.\ starting angle $\phi$ for the GTE family of
    orbits, for four different scaled electric field strengths.
    Action is measured at the orbit's closest approach to the nucleus,
    in units of $\hbar$.  Actual values (symbols) are compared with
    the classical perturbation theory estimate of
    Eq.~(\ref{eq:S-dependence}) (curves).  For $f=0$ (cylindrical
    symmetry), the action is independent of $\phi$.  The two surviving
    orbits occur at the indicated angles $\phi^1$ and $\phi^2 \approx
    \phi^1 + \pi$.  }
  \label{fig:sinusoid}
\end{figure}

The phase of the returning wave from each direction in the
$k^{\text{th}}$ family is given by the classical action of the orbit
returning from that direction.  After symmetry is broken, that action
$S_k^F(\phi)$ (here measured to the orbit's perihelion) depends on
$\phi$, the orbit's azimuthal launching angle, and on $F$, the
electric field strength.  A theorem of classical mechanics asserts
that the change in action caused by the electric field is proportional
to the time-integral of the perturbing Hamiltonian over the {\em
  unperturbed\/} trajectory: $\partial S_k^F/\partial F = - \int
(\partial H/\partial F) dt$.  It follows that, to first order in $F$,
the actions of orbits within a family have a sinusoidal dependence on
$\phi$ \cite{note:ParallelOrbit}:
\begin{eqnarray}
  S_k^F(\phi) & = & S_k^{F=0} + {\bf F} \cdot {\bf d}_k(\phi) \, T_k \\
  & = & S_k^{F=0} + F d_k^\perp T_k \cos(\phi - \phi^1_k) ,
\label{eq:S-dependence}
\end{eqnarray}
(see Fig.~\ref{fig:sinusoid}) where ${\bf d}_k(\phi) \equiv \avg{-{\bf
    r}(t)}$ is the time-averaged dipole moment of the orbit starting
in direction $\phi$, $d_k^\perp$ is the magnitude of the component of
${\bf d}_k$ in the $x$-$y$ plane, and $T_k$ is the return time, all for
the unperturbed orbit.  The two closed orbits that survive the
symmetry breaking are those with the extremal actions; their actions
define the amplitude of the sinusoid.

The returning electron wave near the nucleus is a coherent
superposition of waves arriving from each trajectory's final
direction, each with a phase given by $\exp[i S_k^F(\phi)]$.  The
returning wavefunction for the GTE recurrence, calculated in this way,
is shown in Fig.~\ref{fig:quantum-cusp}(c).

\paragraph{Falloff of recurrence strength}

A detailed analysis shows that the recurrence amplitude for
weakly-broken cylindrical symmetry is the coherent sum of
contributions from each orbit in the family.  Each orbit contributes
nearly the same amplitude, but with a phase that varies according to
Eq.~(\ref{eq:S-dependence}).  The net result is that
Eq.~(\ref{eq:absorption}) is replaced with $\sigma^{F} = C_0 + \sum_k
\sigma_k^F$, where
\begin{eqnarray}
  \sigma_k^F & = & C_k^{F=0} \left\{
      \recip{2\pi} \int_0^{2\pi} d\phi \,
      \sin [ S_k^F(\phi) - \gamma_k ] \right\}
\label{eq:AbsBesselInt} \\
  & = & C_k^{f=0} J_0(f \tilde{d}^\perp_k \tilde{T}_k w) \;
          \sin(\tilde{S}_k^{f=0} w - \gamma_k) .
\label{eq:AbsBessel}
\end{eqnarray}
The integral in Eq.~(\ref{eq:AbsBesselInt}) is over the orbits of family
$k$, parameterized by their initial azimuthal angles $\phi$; i.e., this is
the coherent superposition of waves having phases that vary with $\phi$.
Equation~(\ref{eq:AbsBessel}) follows from substituting
Eq.~(\ref{eq:S-dependence}) into Eq.~(\ref{eq:AbsBesselInt}), evaluating
the integral, and converting to scaled variables.

We see that each oscillation in the absorption spectrum is multiplied
by a Bessel function that accounts for the partially-destructive
interference of the contributing orbits.  The Fourier transform of
Eq.~(\ref{eq:AbsBessel}) over the experimental range of $w$ predicts
the height and shape of each peak in the recurrence spectrum.  This
theoretical prediction, evaluated for the GTE recurrence, is shown in
Fig.~\ref{fig:theory}(b).  It is seen that theory explains all of the
characteristics seen in experiment: the pattern of the falloff, the
location of the minimum, and the location and relative strength of the
revival.

It is to be noted that Eq.~(\ref{eq:AbsBesselInt}) is applicable
through the whole transition---from perfect cylindrical symmetry to
fully-broken symmetry.  In the former case ($f=0$), the integrand in
Eq.~(\ref{eq:AbsBesselInt}) is constant and the recurrence gives only
the usual sinusoidal modulation of the absorption.  In the latter case
($f$ large), the integral can be evaluated by the method of stationary
phase, and the result is the contribution from the two isolated
returning orbits (including correct Maslov indices).  The revival
occurs when their phase difference reaches $2\pi$ and they again
interfere constructively.

\paragraph*{Conclusions.}

The most important result of this work is an illumination of the
difference between classical and quantum mechanical symmetry breaking.
Classically, even the slightest symmetry breaking destroys the perfect
focus, leaving only two surviving closed orbits, and reducing the
density of returning trajectories instantaneously from infinite to
finite.  For strong symmetry breaking, only the two surviving closed
orbits contribute significantly.  However, in quantum mechanics there
is an intermediate regime of weak symmetry breaking for which the
contributions of the whole family of orbits---not just the ones that
remain closed---must be considered together.

In this regime, the strength of the recurrence is determined by the
degree of constructive and destructive interference across the whole
returning wave.  The strong recurrence for exact symmetry corresponds
to pure constructive interference, and the minimum and
revival---purely nonclassical effects---are caused by the dominance of
destructive and constructive interference, respectively, across the
wavefront.  This signature of symmetry breaking is seen in experiment
and is predicted by a uniform semiclassical theory.

We would like to thank J\"org Main for discussions of related quantum
calculations.  The work at the University of Bielefeld was supported
by the Deutsche Forschungs\-ge\-mein\-schaft.  The work at the
College of William and Mary was supported by the Office of Naval
Research and the National Science Foundation.
%This work was supported by DFG, ONR, and NSF.

% References:


\begin{thebibliography}{99}
\bibitem{further-work}
Further details will be presented in a future publication.
\end{thebibliography}
\end{document}